\documentclass[11pt]{article}
\usepackage{latexsym,mathrsfs}
\usepackage{newproof}
\usepackage[thmmarks]{ntheorem}
\usepackage{amsmath}
\usepackage{amssymb}
\usepackage{pstricks,pst-node,pst-tree}
\usepackage{stmaryrd}
\usepackage{pst-3d,pst-coil,pst-poly,ifthen}
\usepackage{url}
\usepackage[dvips]{graphicx}

\theorembodyfont{\rmfamily}
\theoremsymbol{\ensuremath{\dashv}} 
\newtheorem{theorem}{Theorem}
\newtheorem{definition}[theorem]{Definition}
\newtheorem{proposition}[theorem]{Proposition}

\newtheorem{corollary}[theorem]{Corollary}

\newtheorem{example}[theorem]{Example}

\newcommand{\atoms}{P}
\newcommand{\Atoms}{\atoms}
\newcommand{\lang}{{\ensuremath{\mathscr{L}}}}

\newcommand{\agent}{a}

\newcommand{\agents}{A}
\newcommand{\Agents}{\agents}
\newcommand{\group}{B}
\newcommand{\Group}{\group}

\newcommand{\events}{\ensuremath{\mathsf{E}}}

\newcommand{\state}{s}

\newcommand{\stateb}{t}
\newcommand{\statec}{u}

\newcommand{\states}{S}
\newcommand{\States}{\states}

\newcommand{\bewijssysteem}[1]{\ensuremath{\mathbf{#1}}}

\newcommand{\proum}{\ensuremath{\bewijssysteem{UM}}}

\newcommand{\umodel}{\ensuremath{\mathsf{U}}}
\newcommand{\arel}{\ensuremath{\mathsf{R}}}

\newcommand{\isdef}{\ensuremath{=}}
\newcommand{\iffdef}{\mbox{iff}}

\newcommand{\weg}[1]{}

\newcommand{\eq}{\leftrightarrow}

\newcommand{\imp}{\rightarrow}

\newcommand{\et}{\wedge}

\newcommand{\Et}{\bigwedge}
\newcommand{\Vel}{\bigvee}

\newcommand{\T}{\top}
\newcommand{\F}{\bot}

\renewcommand{\phi}{\varphi}
\newcommand{\union}{\cup}

\newcommand{\dia}[1]{\langle #1 \rangle}

\newcommand{\bisim}{\ \underline{\eq}\ }

\newcommand{\x}[1]{\mbox{ #1 }}

\newcommand{\then}{\ ; \ }

\newcommand{\suchthat}{\ | \ }

\newcommand{\pre}{\mathsf{pre}}
\newcommand{\post}{\mathsf{post}}

\newcommand{\Iff}{\mbox{\ iff \ }}

\newcommand{\dom}{\ensuremath{\text{dom}}}

\newcommand{\canonicalvaluation}[1]{%
\ifthenelse{\equal{barteld}{hans}}{%
\ifthenelse{\equal{#1}{}}{\ensuremath{V^c}}{\ensuremath{V^c_{#1}}}}
{\ifthenelse{\equal{#1}{}}{\ensuremath{V^c}}{\ensuremath{V^c(#1)}}}
}
\newcommand{\canonicalrelation}[3]{%
\ifthenelse{\equal{barteld}{hans}}{%
\ifthenelse{\equal{#1}{} \and \equal{#2}{} \and \equal{#3}{}}{\ensuremath{\sim^c}}{%
\ifthenelse{\equal{#1}{} \and
  \equal{#2}{}}{\ensuremath{\sim^c_{#3}}}{\ensuremath{#1 \sim^c_{#3} #2}}}
}{%
\ifthenelse{\equal{#1}{} \and \equal{#2}{} \and \equal{#3}{}}{\ensuremath{R^c}}{%
\ifthenelse{\equal{#1}{} \and
  \equal{#2}{}}{\ensuremath{R^c(#3)}}{\ensuremath{(#1,#2) \in R^c(#3)}}}
}}

\newcommand{\relation}[3]{%
\ifthenelse{\equal{#1}{} \and \equal{#2}{} \and
\equal{#3}{}}{\ensuremath{\sim}}{%
\ifthenelse{\equal{#1}{} \and
\equal{#2}{}}{\ensuremath{\sim_{#3}}}{\ensuremath{#1 \sim_{#3} #2}}}
}

\newcommand{\bnfsep}{\ensuremath{\;\mid\;}}

\newcommand{\groupone}{\ensuremath{\group}}
\newcommand{\powerset}{\ensuremath{\wp}}

\newcommand{\bisrel}{\ensuremath{\mathfrak{R}}}

\newcommand{\eone}{\ensuremath{\mathsf{e}}}
\newcommand{\etwo}{\ensuremath{\mathsf{f}}}
\newcommand{\ethree}{\ensuremath{\mathsf{g}}}

\newcommand{\atom}{p}
\newcommand{\atomb}{q}

\newcommand{\Events}{\events}
\newcommand{\event}{\eone}
\newcommand{\eventb}{\etwo}
\newcommand{\eventc}{\ethree}

\newcommand{\sskip}{\mathsf{skip}}

\newcommand{\eventp}{\mathsf{p}}
\newcommand{\eventnp}{\mathsf{np}}
\newcommand{\eventn}{\mathsf{n}}

\newcommand{\ser}{\ensuremath{\mathsf{ser}}}
\newcommand{\truefalse}{{\T \!\! \F}}
\newcommand{\single}{\ensuremath{\mathsf{one}}}

\usepackage{a4wide}

\title{Semantic results for ontic and epistemic change}
\author{Hans van Ditmarsch\thanks{University of Otago, New Zealand \& IRIT, France, {\tt hans@cs.otago.ac.nz}} \and Barteld Kooi\thanks{University of Groningen, Netherlands, {\tt B.P.Kooi@rug.nl}}}

\begin{document}

\maketitle

\begin{abstract}
We give some semantic results for an epistemic logic incorporating dynamic operators to describe information changing events. Such events include epistemic changes, where agents become more informed about the non-changing state of the world, and ontic changes, wherein the world changes. The events are executed in information states that are modeled as pointed Kripke models. Our contribution consists of three semantic results. (i) Given two information states, there is an event transforming one into the other. The linguistic correspondent to this is that every consistent formula can be made true in every information state by the execution of an event. (ii) A  more technical result is that: every event corresponds to an event in which the postconditions formalizing ontic change are assignments to `true' and `false' only (instead of assignments to arbitrary formulas in the logical language). `Corresponds' means that execution of either event in a given information state results in bisimilar information states. (iii) The third, also technical, result is that every event corresponds to a sequence of events wherein all postconditions are assignments of a single atom only (instead of {\em simultaneous} assignments of more than one atom).
\end{abstract}

\section{Introduction}

In dynamic epistemic logics \cite{plaza:1989,gerbrandyetal:1997,benthem06,baltagetal:2004,hvdetal.del:2007} one does not merely describe the static (knowledge and) beliefs of agents but also dynamic features: how does belief change as a result of events taking place. The main focus of such logics has been change of {\em only} belief, whereas the facts describing the world remain the same. Change of belief is known as {\em epistemic} change. One can also model change of facts, and the resulting consequences of such factual changes for the beliefs of the agents. Change of facts is also known as {\em ontic} change (change of the real world, so to speak).\footnote{In the areas known as `artificial intelligence' and `belief revision', epistemic and ontic change are called, respectively, {\em belief revision} \cite{agm:1985} and {\em belief update} \cite{katsuno:91a}. We will not use that terminology.} In this contribution we use `event' to denote {\em any} sort of information change, both epistemic and ontic. Let us begin by a simple example involving various events.

\begin{example} \label{world.exa.in}
Given are two players Anne and Bill. Anne shakes a cup containing a single coin and deposits the cup upside down on the table (there are no opportunities for cheating). Heads or tails?  Initially, we have a situation wherein both Anne ($a$) and Bill ($b$) are uncertain about the truth of that proposition. A player may observe whether the coin is heads or tails, and/or flip the coin, and with or without the other player noticing that. Four example events are as follows.
\begin{enumerate}
\item Anne lifts the cup and looks at the coin. Bill observes this but is not able to see the coin. All the previous is common knowledge to Anne and Bill. \label{in.one}
\item Anne lifts the cup and looks at the coin without Bill noticing that. \label{in.two}
\item Anne lifts the cup, looks at the coin, and ensures that it is tails (by some sleight of hand). Bill observes Anne looking but is not able to see the coin, and he considers it possible that Anne has flipped the coin to tails (and this is common knowledge). \label{in.three}
\item Bill flips the coin (without seeing it). Anne considers that possible (and this is common knowledge). \label{in.four}
\end{enumerate}
Events 1, 3, and 4 are all public in the sense that the actual event is considered possible by both agents, and that both agents know that, and know that they know that, etc.; whereas event 2 is private: Bill is unaware of the event; the event is private to Anne. Events 3 and 4 involve ontic change, whereas events 1 and 2 only involve epistemic change. Flipping a coin is ontic change: the value of the atomic proposition `the coin is heads' changes from false to true, or from true to false, because of that. But in the case of events 1 and 2 that value, whether true or false, remains unchanged. What changes instead, is how informed the agents are about that value, or about how informed the other agent is. In 1 and 2, Anne still learns whether the coin is head or tails. In 1, Bill `only' learns that Anne has learnt whether the coin is head or tails: he has not gained factual information at all. In Example \ref{examplelater}, later, we will formalize these descriptions.
\end{example}

Various logics have been proposed to model such events. A well-known setting is that of interpreted systems by Halpern et al.\ \cite{faginetal:1995}. Each agent has a local state; the local states of all agents together with a state of the environment form a global state; belief of an agent is modelled as uncertainty to distinguish between global states wherein the agent has the same local state, and change of belief is modelled as a transition from one global state to another one, i.e.\ as a next step in a run through the system. In an interpreted system the treatment of epistemic and ontic change is similar---either way it is just a next step in a run, and how the valuation between different points changes is not essential to define or describe the transition. There is a long tradition in such research \cite{meyden.pricai:1997,faginetal:1995}.

The shorter history of dynamic epistemic logic started by focussing on epistemic change \cite{plaza:1989,gerbrandyetal:1997,benthem06,baltagetal:2004,hvdetal.del:2007}. In that community, how to model ontic change was first mentioned by Baltag, Moss, and Solecki as a possible extension to their {\em action model logic} for epistemic change \cite{baltagetal:1999}. More detailed proposals for ontic change are far more recent \cite{vaneijck:2004,hvdetal.aamas:2005,hvd.pit:2006,jfaketal.lcc:2006,kooi.jancl:2007,renardel:cm,herzigetal:2006,herzigetal:2005b}. The literature will be discussed in more detail in Section \ref{other}.

\bigskip

Section~\ref{three} contains logical preliminaries, including detailed examples. Section~\ref{semanticresults} contains the semantic results that we have achieved for the logic. This is our original contribution to the area. These results are that: $(i)$ for all finite models and for all consistent formulas we can construct an event that `realizes' the formula, i.e., the formula becomes true after execution of the event; that: $(ii)$ every event (with assignments of form $\atom := \phi$, for $\phi$ in the language) corresponds to an event with assignments to true and false only; and also that: $(iii)$ every event corresponds to a sequence of events with assignments for a single atom only. Section~\ref{other} discusses related work in detail. 

\section{A logic of ontic and epistemic change} \label{three}

We separately introduce the logical language, the relevant structures, and the semantics of the language on those structures. The syntax and semantics {\em appear} to overlap: updates are structures that come with preconditions that are formulas. In fact it is properly covered by the double induction used in the language definition, as explained after Definition \ref{updatemodel} below. For a detailed treatment of the logic without ontic change, and many examples, we recommend \cite{hvdetal.del:2007}; for more examples of the logic involving ontic change, see \cite{hvd.pit:2006,hvdetal.aamas:2005,vaneijck:2004}.

\subsection{Language}
We use the style of notation from propositional dynamic logic (PDL) for
modal operators which is also used in \cite{jfaketal.lcc:2006}.
\begin{definition}[Language]
Let a finite set of agents $\Agents$ and a countable set of propositional variables $\Atoms$ be given. The language $\lang$ is given by the following BNF:
\[
\begin{array}{lll}
\phi &::=& \atom \bnfsep \neg \phi \bnfsep \phi \et \phi \bnfsep [\alpha]\phi \\
\alpha &::=& \agent \bnfsep \Group^* \bnfsep (\umodel,\eone)
\end{array}
\]
where $\atom\in\Atoms$, $\agent\in\Agents$, $\Group \subseteq \Agents$ (the dynamic operator $[\Group^*]$ is associated with `common knowledge among the agents in $\Group$'), and where $(\umodel,\eone)$ is an {\em update} as (simultaneously) defined below.
\end{definition}

We use the usual abbreviations, and conventions for deleting parentheses. In particular, $[\Group]\phi$ stands for $\Et_{\agent\in\Group} [\agent] \phi$, and (the diamond form) $\dia{\alpha} \phi$ is equivalent to $\neg [\alpha] \neg \phi$. Non-deterministic updates are introduced by abbreviation: $[(\umodel,\eone) \union (\umodel',\etwo)]\phi$ is by definition $[\umodel,\eone]\phi \et [\umodel',\etwo]\phi$.

\subsection{Structures}

\paragraph*{Epistemic model} The models which adequately present an information state in a multi-agent environment are Kripke models from epistemic logic. The set of states together with the accessibility relations represent the information the agents have. If one state $\state$ has access to another state $\stateb$ for an agent $\agent$, this means that, if the actual situation is $\state$, then according to $\agent$'s information it is possible that $\stateb$ is the actual situation.

\begin{definition}[Epistemic model]
Let a finite set of agents $\Agents$ and a countable set of propositional variables $\Atoms$ be given. An epistemic model is a triple $M=(\States,R,V)$ such that
\begin{itemize}
\item {\em domain} $\States$ is a non-empty set of possible states,
\item $R:\Agents \rightarrow \powerset (\States \times \States)$ assigns an {\em accessibility relation} to each agent $\agent$,
\item $V:\Atoms \rightarrow \powerset(\States)$ assigns a set of states to each propositional variable; this is the {\em valuation} of that variable.
\end{itemize}
A pair $(\States,R)$ is called an {\em epistemic frame}. A pair $(M,\state)$, with $\state \in \States$, is called an {\em epistemic state}.\weg{\footnote{`Epistemic state' in our sense is not to be confused with the set of worlds/states indistinguishable for a given agent from a given state, i.e., $R(\agent)(\state)\isdef \{ \stateb \suchthat (\state,\stateb)\in R(\agent) \}$. This set $R(\agent)(\state)$ is also commonly called an epistemic state. We call that the set of reachable states, or, in the case of $KD45$ or $S5$-agents, the epistemic class.}}
\end{definition}

A well-known notion of sameness of epistemic models is `bisimulation'. Several of our results produce models that are bisimilar: they correspond in the sense that even when not identical (isomorphic), they still cannot be distinguished in the language.
\begin{definition}[Bisimulation] \label{bisim}
Let two models $M= (\States ,R,V)$ and $M'= (\States',R',V')$ be given. A non-empty relation $\bisrel \subseteq \States
\times \States'$ is a bisimulation iff for all $\state
\in \States$ and $\state' \in \States'$ with $(\state,\state') \in
\bisrel$:
\begin{description}
\item[atoms] for all $p \in \atoms$: $\state \in V(p)$ iff $\state' \in V'(p)$; 
\item[forth] for all $\agent \in \Agents$ and all $\stateb \in \States$: if
$(\state,\stateb) \in R(\agent)$, then there is a $\stateb'\in
\States'$ such that $(\state',\stateb') \in R'(\agent)$ and
$(\stateb,\stateb') \in \bisrel$;
\item[back] for all $\agent \in \agents$ and all $\stateb' \in \States'$:
if $(\state',\stateb') \in R'(\agent)$, then there is a
$\stateb \in \States$ such that $(\state,\stateb) \in R(\agent)$
and $(\stateb,\stateb') \in \bisrel$.
\end{description}
\end{definition}
We write $(M,\state) \bisim (M',\state')$, iff there is a bisimulation
between $M$ and $M'$ linking $\state$ and $\state'$, and we then call
$(M,\state)$ and $(M',\state')$ bisimilar. A model such that all bisimilar states are identical is called a {\em bisimulation contraction} (also known as a {\em strongly extensional model}).

\paragraph*{Update model\footnote{In the literature update models are also
called action models. Here we follow \cite{jfaketal.lcc:2006} and call them
update models, since no agency seems to be involved.}} An epistemic model
represents the information of the agents. \emph{Information change} is
modelled as changes of such a model. There are three variables. One can
change the set of states, the accessibility relations and the valuation. It
may be difficult to find the exact change of these variables that matches a
certain description of an information changing event. It is often easier to
think of such an event separately. One can model an information changing
event in the same way as an information state, namely as some kind of
Kripke model: there are various possible events, which the agents may not
be able to distinguish. This is the domain of the model. Rather than a
valuation, a {\em precondition} captures the conditions under which such
events may occur, and {\em postconditions} also determine what epistemic
models may evolve into. Such a Kripke model for events is called an {\em
update model}, which were first studied by Baltag, Moss and Solecki, and
extended with simultaneous substitutions by van Eijck
\cite{baltagetal:1999,vaneijck:2004}. Here we use van Eijck's definition.

\begin{definition}[Update model] \label{updatemodel}
An {\em update model} for a finite set of agents $\Agents$ and a language $\lang$ is a quadruple $\umodel = (\events,\arel,\pre,\post)$ where
\begin{itemize}
\item {\em domain} $\events$ is a finite non-empty set of events,
\item $\arel : \Agents \to \powerset (\events \times \events)$ assigns an {\em accessibility relation} to each agent,
\item $\pre: \events \to \lang$ assigns to each event a {\em precondition},
\item $\post: \events \to (\Atoms \to \lang)$ assigns to each event a {\em postcondition} for each atom. Each $\post(\eone)$ is required to be only finitely different from the identity $\mathsf{id}$; the finite difference is called its {\em domain} $\dom(\post(\eone))$.
\end{itemize}
A pair $(\umodel,\eone)$ with a distinguished actual event $\eone \in
\events$ is called an {\em update}. A pair $(\umodel, \events)$ with
$\events'\subseteq\events$ and $|\events'| > 1$ is a {\em multi-pointed
update}, first introduced in \cite{vaneijck.demo:2004}. The event $\eone$
with $\pre(\eone) = \T$ and $\post(\eone) = \mathsf{id}$ we name
$\sskip$. An update with a singleton domain, accessible to all agents, and
precondition $\T$, is a {\em public assignment}. An update with a singleton
domain, accessible to all agents, and identity postcondition, is a {\em
public announcement}.
\end{definition}
Instead of \begin{quote} {\em $\pre(\eone) = \phi$ and $\post(\eone)(p_1) = \psi_1$, ..., and $\post(\eone)(p_n) = \psi_n$}
\end{quote} we also write\footnote{The notation is reminiscent of that for a {\em knowledge-based program} in the interpreted systems tradition. We discuss the correspondence in Section \ref{other}.} \begin{quote} {\em for event $\eone$: if $\phi$, then $p_1 := \psi_1$, ..., and $p_n := \psi_n$.} \end{quote}
The event $\sskip$ stands for: nothing happens except a tick of the clock.

To see an update as part of the language we observe that: an update
$(\umodel,\event)$ is an inductive construct of type $\alpha$ that is built
the frame underlying $\umodel$ (we can assume a set enumerating such
frames) and from simpler constructs of type $\phi$, namely the
preconditions and postconditions for the events of which the update
consists. This means that there should be a finite number of preconditions
and a finite number of postconditions only, otherwise the update would be
an infinitary construct. A finite number of preconditions is guaranteed by
restricting ourselves in the language to {\em finite} update models. A
finite number of postconditions is guaranteed by (as well) restricting
ourselves to {\em finite} domain for postconditions. this situation is
similar to the case of automata-PDL \cite[chapter~10,
section~3]{hareletal:2000}.

If in case of nondetermistic updates the underlying models are the same, we can also see this as executing a multi-pointed update. For example, $(\umodel,\eone) \union (\umodel,\etwo)$ can be equated with $(\umodel, \{\eone,\etwo\})$.

\begin{example} \label{examplelater}
Consider again the scenario of Example \ref{world.exa.in} on page \pageref{world.exa.in}. Let atomic proposition $\atom$ stand for `the coin lands heads'. The initial information state is represented by a two-state epistemic model with domain $\{1,0\}$, with universal access for $a$ and $b$, and with $V(\atom) = \{1\}$. We further assume that the actual state is $1$. This epistemic state is depicted in the top-left corner of Figure \ref{world.figtails}. The events in Example \ref{world.exa.in} can be visualized as the following updates. The actual state is underlined.
\begin{enumerate}
\item Anne lifts the cup and looks at the coin. Bill observes this but is not able to see the coin. All the previous is common knowledge to Anne and Bill. 
\[
\setlength{\arraycolsep}{0.8cm}
\begin{array}{ll}
& \\
\Rnode{h}{\underline{\eventp}} & \Rnode{t}{\eventnp} \\
&
\end{array}
\psset{nodesep=5pt, labelsep=3pt}
\ncline{<->}{h}{t} \ncput*{b}
\nccircle[angle=90]{->}{h}{.5} \ncput*{a,b}
\nccircle[angle=270]{->}{t}{.5} \ncput*{a,b}
\]
Here, $\pre(\eventp) = \atom$, $\post(\eventp) = \mathsf{id}$, $\pre(\eventnp) = \neg \atom$, $\post(\eventnp) = \mathsf{id}$. The update model consists of two events. The event $\eventp$ corresponds to Anne seeing heads, and the event $\eventnp$ to Anne seeing tails; Anne is aware of that: thus the reflexive arrows (identity relation). Bill cannot distinguish them from one another: thus the universal relation. The aspect of common knowledge, or common awareness, is also present in this dynamic structure: the reflexive arrow for Anne also encodes that Anne knows that she lifts the cup and that Bill observes that; similarly for Bill, and for iterations of either awareness.
\item Anne lifts the cup and looks at the coin without Bill noticing that. 
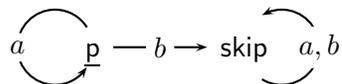
\[
\setlength{\arraycolsep}{0.8cm}
\begin{array}{ll}
& \\
\Rnode{h}{\underline{\eventp}} & \Rnode{t}{\sskip} \\
&
\end{array}
\psset{nodesep=5pt, labelsep=3pt}
\ncline{->}{h}{t} \ncput*{b}
\nccircle[angle=90]{->}{h}{.5} \ncput*{a}
\nccircle[angle=270]{->}{t}{.5} \ncput*{a,b}
\]
Event $\eventp$ is as in the previous item and $\sskip$ is as above. In this update, there is no common awareness of what is going on: Anne observes heads knowing that Bill is unaware of that, whereas Bill does not consider the actual event; the $b$-arrow points to the other event only. 
\item Anne lifts the cup, looks at the coin, and ensures that it is tails (by some sleight of hand). Bill observes Anne looking but is not able to see the coin, and he considers it possible that Anne has flipped the coin to tails (and this is common knowledge).
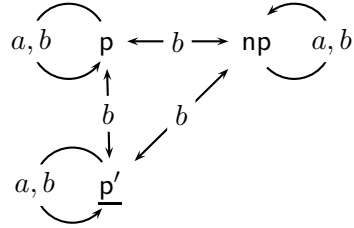
\[
\setlength{\arraycolsep}{0.8cm}
\begin{array}{ll}
& \\
\Rnode{ht}{\eventp} & \Rnode{tt}{\eventnp} \\ \ & \ \\ \ & \ \\ \ & \ \\
\Rnode{hb}{\underline{\eventp'}} & \\
\end{array}
\psset{nodesep=5pt, labelsep=3pt}
\ncline{<->}{ht}{tt} \ncput*{b}
\ncline{<->}{hb}{ht} \ncput*{b}
\ncline{<->}{tt}{hb} \ncput*{b}
\nccircle[angle=90]{->}{ht}{.5} \ncput*{a,b}
\nccircle[angle=270]{->}{tt}{.5} \ncput*{a,b}
\nccircle[angle=90]{->}{hb}{.5} \ncput*{a,b}
\]
Events $\eventp$ and $\eventnp$ are as before, whereas $\pre(\eventp') = \T$, $\post(\eventp')(\atom) = \F$. The event $\eventp'$ may take place both when the coin is heads and when the coin is tails, in the first case atom $\atom$ is set to false (tails), and in the second it remains false.
\item Bill flips the coin (without seeing it). Anne considers that possible (and this is common knowledge).
\[
\setlength{\arraycolsep}{0.8cm}
\begin{array}{ll}
& \\
\Rnode{h}{\underline{\eventn}} & \Rnode{t}{\sskip} \\
&
\end{array}
\psset{nodesep=5pt, labelsep=3pt}
\ncline{<->}{h}{t} \ncput*{a}
\nccircle[angle=90]{->}{h}{.5} \ncput*{a,b}
\nccircle[angle=270]{->}{t}{.5} \ncput*{a,b}
\]
Here, $\pre(\eventn) = \T$, $\post(\eventn)(\atom) = \neg \atom$, and $\sskip$ is as before. For models where all accessibility relations are equivalence relations we will also use a simplified visualization that merely links states in the same equivalence class. E.g., this final event is also depicted as:
\[
\setlength{\arraycolsep}{0.8cm}
\begin{array}{ll}
\Rnode{h}{\underline{\eventn}} & \Rnode{t}{\sskip}
\end{array}
\psset{nodesep=5pt, labelsep=3pt}
\ncline{h}{t} \ncput*{a}
\]
\end{enumerate}
\end{example}

\subsection{Semantics}

The semantics of this language is standard for epistemic logic and based on the product construction for the execution of update models from the previous section. Below, $R(\Group)^*$ is the transitive and reflexive closure of the union of all accessibility relations $R(\agent)$ for agents $\agent\in\Group$. Definitions \ref{sema} and \ref{semb} are supposed to be defined simultaneously.

\begin{definition}[Semantics] \label{sema}
Let a model $(M,\state)$ with $M=(\States,R,V)$ be given. Let $\agent \in \Agents$, $\Group \subseteq \Agents$, and $\phi, \psi \in \lang$.
\[
\begin{array}{lcl}
(M,\state) \models \atom  & \text{iff} & \state \in V(\atom)\\
(M,\state) \models \neg \phi  & \text{iff} &  (M,\state) \not \models \phi\\
(M,\state) \models \phi \et \psi  & \text{iff} &  (M,\state) \models \phi \text{ and   } (M,\state) \models \psi\\
(M,\state) \models [\agent]\phi  & \text{iff} &  (M,\stateb) \models \phi
\text{ for all $\stateb$ such that } (\state,\stateb) \in R(\agent)\\
(M,\state) \models [\Group^*]\phi & \text{iff} &  (M,\stateb) \models \phi \text{ for all $\stateb$ such that } (\state,\stateb) \in R(\Group)^*\\
(M,\state) \models [\umodel,\eone]\phi & \text{iff}& (M,\state)\models \pre(\eone) \text{ implies }
(M \otimes \umodel, (\state,\eone)) \models \phi \\
\end{array}
\]
\end{definition}

\begin{figure}
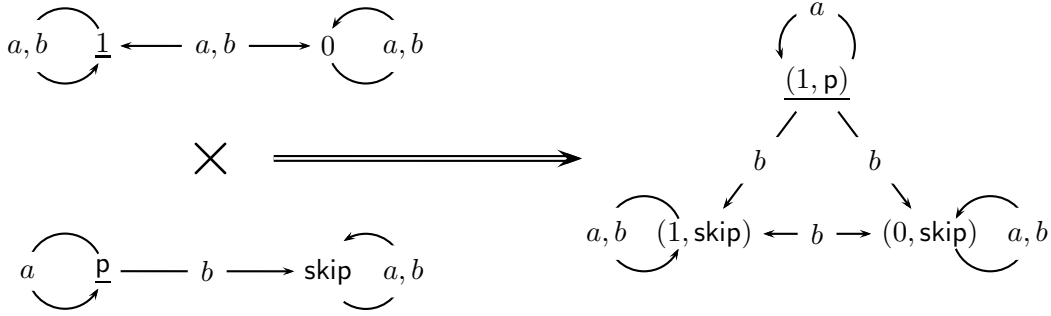

\psset{border=2pt, nodesep=4pt, radius=2pt, tnpos=a}
\pspicture(0,.5)(13,5.5)
\rput(3,4.5){\rnode{top}{
\pspicture(0,-0.5)(5,.5)
\rput(1,0){\rnode{1}{$\underline{1}$}}
\rput(4,0){\rnode{0}{$0$}}
\ncline{<->}{1}{0} \ncput*{$a,b$}
\nccircle[angle=90]{->}{1}{.5} \ncput*{$a,b$}
\nccircle[angle=270]{->}{0}{.5} \ncput*{$a,b$}
\endpspicture
}}
\rput(3,3){\rnode{middle}{\Huge $\times$}}
\rput(3,1.5){\rnode{left}{
\pspicture(0,-0.5)(5,.5)
\rput(1,0){\rnode{p}{$\underline{\eventp}$}}
\rput(4,0){\rnode{skip}{$\sskip$}}
\ncline{->}{p}{skip} \ncput*{$b$}
\nccircle[angle=90]{->}{p}{.5} \ncput*{$a$}
\nccircle[angle=270]{->}{skip}{.5} \ncput*{$a,b$}
\endpspicture
}}
\rput(11,3){\rnode{right}{
\pspicture(0,-0.5)(5,2.5)
\rput(2.5,2){\rnode{1p}{$\underline{(1,\eventp)}$}}
\rput(.7,0){\rnode{h1skip}{$ \ $}}
\rput(4.3,0){\rnode{h0skip}{$ \ $}}
\rput(1,0){\rnode{1skip}{$(1,\sskip)$}}
\rput(4,0){\rnode{0skip}{$(0,\sskip)$}}
\ncline{->}{1p}{1skip} \ncput*{$b$}
\ncline{->}{1p}{0skip} \ncput*{$b$}
\ncline{<->}{1skip}{0skip} \ncput*{$b$}
\nccircle{->}{1p}{.5} \ncput*{$a$}
\nccircle[angle=90]{->}{h1skip}{.5} \ncput*{$a,b$}
\nccircle[angle=270]{->}{h0skip}{.5} \ncput*{$a,b$}
\endpspicture
}}
\ncline[nodesep=.5, doubleline=true, arrows=->]{middle}{right}
\endpspicture
\caption{In an epistemic state where Anne and Bill are uncertain about the truth of $\atom$ (heads or tails), and wherein $\atom$ is true, Anne looks at the coin without Bill noticing it. The actual states and events are underlined.}
\label{world.figtails}
\end{figure}

We now define the effect of an update on an epistemic state---Figure \ref{world.figtails} gives an example of such update execution.

\begin{definition}[Execution] \label{semb}
Given are an epistemic model $M = (\States,R,V)$, a state $\state \in \States$, an update model $\umodel = (\events, \arel, \pre, \post)$, and an event $\eone \in \events$ with $(M,\state) \models\pre(\eone)$. The result of executing $(\umodel,\eone)$ in $(M,\state)$ is the model $(M \otimes \umodel, (\state,\eone))=((\States',R',V'),(\state,\eone))$ where
\begin{itemize}
\item $\States'=\{(\stateb,\etwo) \mid (M,\stateb)\models \pre(\etwo)\}$,
\item $R'(\agent)=\{((\stateb,\etwo),(\statec,\ethree)) \mid (\stateb,\etwo),(\statec,\ethree) \in \States' \text{ and } (\stateb,\statec) \in R(\agent) \text{ and } (\etwo,\ethree) \in \arel(\agent)\}$,
\item $V'(\atom)=\{(\stateb,\etwo) \mid (M,\stateb) \models \post(\etwo)(\atom) \}$.
\end{itemize}
\end{definition}

\begin{definition}[Composition of update models] \label{composition}
Let update models $\umodel=(\events,\arel,\pre,\post)$ and $\umodel'=(\events',\arel',\pre',\post')$ and events $\eone \in \events$ and $\eone' \in \events'$ be given. The composition $(\umodel,\eone) \then (\umodel',\eone')$ of these update models is  $(\umodel'',\eone'')$ where $\umodel''=(\events'',\arel'',\pre'',\post'')$ is defined as
\begin{itemize}
\item $\events''=\events \times \events'$,
\item $\arel''(\agent)=\{((\etwo,\etwo'),(\ethree,\ethree')) \mid (\etwo,\ethree) \in \arel(\agent) \text{
and } (\etwo',\ethree') \in \arel'(\agent)\}$,
\item $\pre''(\etwo,\etwo')= \pre(\etwo) \et [\umodel,\etwo]\pre'(\etwo')$,
\item $\dom(\post''(\etwo,\etwo'))=\dom(\post(\etwo)) \cup
  \dom(\post'(\etwo'))$ and if $\atom \in \dom(\post''(\etwo,\etwo'))$, then
\[\post''(\etwo,\etwo')(\atom) = \left\{ \begin{array}{ll}
  \post(\etwo)(\atom) & \text{if } \atom \not \in \dom(\post'(\etwo')),\\
  {[}\umodel,\etwo]\post'(\etwo')(\atom) & \text{otherwise.}\\
                      \end{array}
\right.
\]
\end{itemize}
\end{definition}
The reason for ${[}\umodel,\etwo]\post'(\etwo')(\atom)$ in the final clause will become clear from the proof detail shown for Proposition \ref{compositionprop}.

\begin{proposition} \label{compositionprop}
$\models [\umodel,\eone][\umodel',\eone']\phi \leftrightarrow [(\umodel,\eone) \then (\umodel',\eone')]\phi$
\end{proposition}

\begin{proof}
Let $(M,\stateb)$ be arbitrary. To show that $(M,\stateb) \models [(\umodel,\event) \then (\umodel',\event')] \phi$ if and only if $(M,\stateb) \models [\umodel,\event][\umodel',\event'] \phi$, it suffices to show that $M \otimes (\umodel \then \umodel')$ is isomorphic to $(M \otimes \umodel) \otimes \umodel')$. A detailed proof (for purely epistemic updates) is found in \cite{hvdetal.del:2007}. The postconditions (only) play a part in the proof that the valuations correspond:

For the valuation of facts $\atom$ in the domain of $\post''$ we distinguish the cases ( \ $\atom\in\dom(\post(\event))$ but $\atom\not\in\dom(\post'(\event'))$ \ ) $(i)$, and otherwise $(ii)$. The valuation of a $i$-atom in a triple $(\stateb, (\event,\event'))$ is $\post(\event)(\atom)$ according to the definition of updates composition; and the valuation of a $ii$-atom is $[\umodel,\event]\post'(\event')(\atom)$. Consider the corresponding triple $((\stateb,\event),\event')$. The valuation of an $i$-atom in $(\stateb,\event)$ is $\post(\event)(\atom)$, and because $\atom$ does not occur in $\dom(\post'(\event'))$ its value in the triple $((\stateb,\event),\event')$ will remain the same. For a $ii$-atom, its final value will be determined by evaluating $\post'(\event')(\atom)$ in $((M \otimes \umodel), (\stateb,\event))$. This corresponds to evaluating $[\umodel, \event]\post'(\event')(\atom)$ in $(M,\stateb)$.
\end{proof}

\subsection{Proof system}

A proof system $\proum$ for the logic is given in Table~\ref{ps}. The proof
system is a lot like the proof system for the logic of epistemic actions (i.e., for the logic {\em without} postconditions to model valuation change) in
\cite{baltagetal:1999}. There are two differences. The axiom `atomic
permanence' in \cite{baltagetal:1999}---$[\umodel,\eone]\atom \eq
(\pre(\eone) \imp \atom)$---is now instead an axiom expressing when atoms
are {\em not} permanent, namely how the value of an atom can change,
according to the postcondition for that atom: \[ [\umodel,\eone]\atom \eq
(\pre(\eone) \imp \post(\eone)(\atom)) \hspace{2cm} \hfill \text{update and
atoms} \] The second difference is not apparent from Table~\ref{ps}. The
axiom \[ [\umodel,\eone] [\umodel',\eone'] \phi \eq [(\umodel,\eone) \then
(\umodel',\eone')] \phi \hspace{2cm} \hfill \text{update composition} \]
also occurs in \cite{baltagetal:1999}. But
Definition~\ref{composition} to compute that composition is in our case a more complex
construction than the composition of update models with only
preconditions, because it also involves resetting the postconditions. We find it remarkable that these are the only differences:
the interaction between {\em postconditions} for an atom and the logical
operators, {\em only} occurs in the axiom where that atom is mentioned, or
implicitly, whereas the interaction between {\em preconditions} and the
logical operators appears in several axioms and rules.
\begin{table}
\framebox{
\begin{tabular}{ll}
all instantiations of propositional tautologies \\
$[\alpha] (\phi \imp \psi) \imp ([\alpha] \phi
   \imp [\alpha] \psi)$ & distribution\\ 
From $\phi$ and $\phi \imp \psi$, infer $\psi$ & modus ponens\\
From $\phi$, infer $[\alpha] \phi$ & necessitation\\ 
$[\umodel,\eone]\atom \eq (\pre(\eone) \imp \post(\eone)(\atom))$ & update and atoms\\
$[\umodel,\eone] \neg \phi \eq (\pre(\eone) \imp \neg [\umodel,\eone]
   \phi)$ & update and negation\\
$[\umodel,\eone] (\phi \et \psi) \eq ([\umodel,\eone]\phi \et [\umodel,\eone]\psi)$ &
   update and conjunction \\
$[\umodel,\eone] [\agent] \phi \eq (\pre(\eone) \imp
   \Et_{(\eone,\etwo) \in \arel(\agent)} [\agent]
    [\umodel,\etwo] \phi)$ & update and knowledge \\
$[\umodel,\eone] [\umodel',\eone'] \phi \eq [(\umodel,\eone);(\umodel',\eone')] \phi$ &
    update composition\\
$[\groupone^*] \phi \imp (\phi \et
   [\groupone][\groupone^*] \phi)$ & mix\\
$[\groupone^*] (\phi \imp [\groupone] \phi)
   \imp (\phi \imp [\groupone^*] \phi)$ & induction axiom \\ & \\
\parbox[t]{7cm}{ Let $(\umodel,\eone)$ be an update model and let a set of
  formulas $\chi_\etwo$ for every $\etwo$ such that $(\eone,\etwo) \in
  \arel(\groupone)^*$ be given.  From $\chi_\etwo \imp [\umodel,\etwo]\phi$
  and $(\chi_\etwo \et \pre(\etwo)) \imp [\agent] \chi_\ethree$ for every
  $\etwo \in \events$ such that $(\eone,\etwo) \in \arel(\groupone)^*$,
  $\agent \in \groupone$ and $(\etwo,\ethree) \in \arel(\agent)$, infer
  $\chi_\eone \imp [\umodel,\eone][\groupone^*] \phi$.} &
\parbox[t]{5cm}{\raggedright updates and common knowledge}
  \\
\end{tabular}}
\caption{The proof system $\proum$.} \label{ps}
\end{table}

The proof system is sound and complete. The soundness of the `update and
atoms' axiom is evident. The soundness of the `update composition' axiom was established in Proposition \ref{compositionprop}. We proved completeness of the logic as a modification of the completeness proof for the logic without ontic change---action model logic---as found in \cite{hvdetal.del:2007}, which in turn is a simplified version of the original proof for that logic as found in \cite{baltagetal:1999}. We do not consider the modified proof of sufficient original interest to report on in detail.

\section{Semantic results} \label{semanticresults}

We now present some semantic peculiarities of the logic. These we deem our contribution to the area. The results help to relate different approaches combining ontic and epistemic change (see Section \ref{other}). The various `normal forms' for update models that we define are also intended to facilitate future tool development. Finally, they are relevant when modelling AGM belief revision \cite{agm:1985} in a dynamic epistemic setting.

\subsection{Arbitrary belief change}

Let $(M,\state)$ and $(M', \state')$ be arbitrary finite epistemic states for the same set of atoms and agents, with $M = (\States, R, V )$ and $M' = (\States', R', V' )$. Surprisingly enough, there is almost always an update transforming the former into the latter. There are two restrictions. Both restrictions are technical and not conceptual. Firstly, for the set of agents with non-empty access in $M'$ there must be a submodel of $M$ containing actual state $\state$ that is serial for those agents. In other words, if an agent initially has empty access and therefore believes everything (`is crazy') you cannot change his beliefs, but otherwise you can. This seems reasonable. Secondly, models $M$ and $M'$ should only differ in the value of a finite number of atoms; more precisely, if we define that \begin{quote} an atom is {\em relevant} in a model iff its valuation is neither empty nor the entire domain, \end{quote} then the requirement is that only a finite number of atoms is relevant in $M \union M'$. This is required, because we can only change the value of a finite number of atoms in the postconditions. This also seems reasonable: as both models are finite, the agents can only be uncertain about the value of a {\em finite} number of atoms (in the combined models $M$ and $M'$); in other words, they are `not interested' in the value of the remaining atoms.

For expository purposes we initially assume that all agents consider the actual state $\state$ of $M$ a possibility (as in all $S5$ models, such as Kripke models representing interpreted systems), thus satisfying the first of the two restrictions above: the serial submodel required is then the singleton model consisting of $\state$, accessible to all agents. The update transforming $(M,\state)$ into $(M',\state')$ can be seen as the composition of two intuitively more appealing updates. That will make clear how we can also describe the required update in one stroke.

In the first step we get rid of the structure of $M$. As the epistemic state $(M,\state)$ is finite, it has a characteristic formula $\delta_{(M,\state)}$ \cite{barwiseetal:1996,jfak.odds:1998}.\footnote{A characteristic formula $\phi$  for a state $(M,\state)$ satisfies that for all $\psi$, $(M,\state) \models \phi$ iff $\phi \models \psi$. In fact, for the construction we only need formulas that can distinguish all states in the domain from one another, modulo bisimilarity. Characteristic formulas satisfy that requirement.} We let the agents publicly learn that characteristic formula.  This event is represented by the singleton update $(\umodel,\event)$ defined as \[ ((\{ \event \}, \arel, \pre, \post), \event) \text{ with } \pre(\event) = \delta_{(M,\state)}, \text{ for all } \agent: (\event,\event) \in \arel(\agent), \text{ and } \post(\event) = \mathsf{id} \] In other words, the structure of the current epistemic state is being publicly announced. The resulting epistemic state is, of course, also singleton, or bisimilar to a singleton epistemic state, as $\delta_{(M,\state)}$ holds in all states in $M$ bisimilar to $\state$. Without loss of generality assume that it is singleton. Its domain is $\{(\state,\event)\}$. This pair $(\state,\event)$ is accessible to itself because for all agents, $(\state,\state) \in R(\agent)$ (all agents consider the actual state $\state$ a possibility), and $(\event,\event) \in \arel(\agent)$. The valuation of propositional variables in this intermediate state are those of state $\state$ in $M$. What the value is does not matter: we will not use that valuation.

Now proceed to the second step. In the epistemic state wherein the agents have common knowledge of the facts in $\state$, the agents learn the structure of the resulting epistemic state $M' = (\States', R', V')$ and their part in it by executing update $(\umodel',\state')$ defined as \[ ((\States', R', \pre', \post'),\state') \text{ with for } \stateb' \in \States':  \pre'(\stateb') = \T \text{ and for relevant } \atom: \post'(\stateb')(\atom) = \T \text{ iff } \stateb' \in V'(\atom) \] Note that the domain $\States'$ and the accessibility relation $R'$ of $\umodel'$ are precisely those of $M'$, the resulting final epistemic model. The postcondition $\post'$ is well-defined, as only a finite number of atoms (the relevant atoms) is considered. Because we execute this update in a singleton model with public access, and because it is executable for every event $\stateb'$, the resulting epistemic state has the same structure as the update: it returns $\States'$ and $R'$ again. The postcondition delivers the required valuation of atoms in the final model: for each {\em event} $\stateb'$ in $\umodel'$ and for all revelant atoms $\atom$, $\atom$ become true in $\stateb'$ if $\atom$ is true in {\em state} $\stateb'$ in $M'$ ($\post(\stateb')(\atom)=\T$), else $\atom$ becomes false. The value of irrelevant atoms remains the same.

We combine these two updates into one by requiring the precondition of the first and the postcondition of the second. Consider $\umodel'_r$ that is exactly as $\umodel'$ except that in all events $\stateb'$ in its domain the precondition is not $\T$ but $\delta_{(M,\state)}$: the characteristic formula of {\em the point} $\state$ of $(M,\state)$. Update $(\umodel'_r,\state')$ does the job: epistemic state $(M \otimes \umodel'_r, (\state,\state'))$ is isomorphic to $(M',\state')$. This will be Corollary \ref{isom} of our more general result, to follow.

\bigskip 

Now consider the more general case that the agents with non-empty access in $M'$ are serial in a submodel $M^\ser$ of $M$ that contains $\state$, with domain $\States^\ser$. In other words: at least all agents who finally have consistent beliefs in some states, initially have consistent beliefs in all states. The construction above will no longer work: if the actual state is not considered possible by an agent, then that agent has empty access in actual state $(\state,\state')$ of $(M \otimes \umodel'_r)$, but not in $\state'$ in $M'$. But if we relax the precondition $\delta_{(M,\state)}$, for the point $\state$ of $(M,\state)$, to the disjunction $\Vel_{\statec\in\States^\ser} \delta_{(M,\statec)}$, that will now carry along the serial subdomain $\States^\ser$, the construction will work because an agent can then always imagine {\em some} state wherein the update has been executed, even it that is not the actual state. This indeed completes the construction.
\begin{definition}[Update for arbitrary change] \label{achange}
Given finite epistemic models $M = (\States, R, V )$ and $M' = (\States', R', V' )$ for the same sets of agents and atoms. Assume that all agents with non-empty access in $M'$ are serial in $M^\ser$ containing $\state$. The update for arbitrary change $(\umodel'', (\state,\state')) = ((\events'', \arel'', \pre'', \post''), (\state,\state'))$ is defined as (for arbitrary agents $\agent$ and arbitrary relevant atoms $\atom$):
\[ \begin{array}{lcl}
\events'' & \isdef & \States' \\
(\stateb',\statec') \in \arel''(\agent) & \iffdef & (\stateb',\statec') \in R'(\agent) \\
\pre''(\stateb') & \isdef & \Vel_{\statec\in\States^\ser} \delta_{(M,\statec)} \\
\post''(\stateb')(\atom) & \isdef & \left\{ \begin{array}{ll}\T & \x{if} \stateb' \in V'(\atom) \\
 \F & \x{otherwise}
                         \end{array}
\right. \\
\end{array} \]
\end{definition}
The epistemic state $(M \otimes \umodel'', (\state,\state'))$ is bisimilar to the stipulated $(M',\state')$, which is the desired result. It will typically not be isomorphic: $M \otimes \umodel''$ can be seen as consisting of a number of copies of $M'$ (namely $|\States^\ser|$ copies) `with the accessibility relations just right to establish the bisimulation'. One copy may not be enough, namely when the state $\stateb$ in $M$ to which that copy corresponds, lacks access for some agents. This access will then also be `missing' between the states of $(\{\stateb\} \times \States')$. But because of seriality one of the other $M'$ copies will now make up for this lack: there is a $\statec\in\States^\ser$ such that $(\stateb,\statec) \in R(\agent)$, which will establish access when required, as in the proof of the following proposition.

\begin{proposition} \label{arbit}
Given $(M,\state)$, $(M', \state')$, and $\umodel''$ as in Definition \ref{achange}. Then ${\mathfrak R}: ((M \otimes \umodel''), (\state,\state')) \bisim (M',\state')$ by way of, for all $\stateb\in\States^\ser$ and $\stateb'\in\States'$: ${\mathfrak R}(\stateb, \stateb') = \stateb'$.
\end{proposition}

\begin{proof} Let $R^\otimes$ be the accessibility relation and $V^\otimes$ the valuation in $(M \otimes \umodel'')$.

{\bf atoms}: \ \
For an arbitrary relevant atom $\atom$: $(\stateb,\stateb') \in V^\otimes(\atom)$ iff $(M,\stateb) \models \post''(\stateb')(\atom)$, and by definition of $\post''$ we also have that $(M,\stateb) \models \post''(\stateb')(\atom)$ iff $\stateb'\in V'(\atom)$. Irrelevant atoms do not change value.

{\bf forth}: \ \ Let $((\stateb_1, \stateb'_1), (\stateb_2, \stateb'_2)) \in R^\otimes(\agent)$  and $((\stateb_1,\stateb'_1), \stateb'_1) \in {\mathfrak R}$. From $((\stateb_1, \stateb'_1), (\stateb_2,\stateb'_2)) \in R^\otimes(\agent)$ follows $(\stateb'_1,\stateb'_2) \in R'(\agent)$. By definition of ${\mathfrak R}$ we also have $((\stateb_2,\stateb'_2), \stateb'_2) \in {\mathfrak R}$.

{\bf back}: Let $((\stateb_1,\stateb'_1), \stateb'_1) \in {\mathfrak R}$ and $(\stateb'_1,\stateb'_2) \in R'(\agent)$. As $M^\ser$ is serial for $\agent$, and $\stateb_1 \in \States^\ser$, there must be a $\stateb_2$ such that $(\stateb_1,\stateb_2)\in R(\agent)$. As $(M,\stateb_2) \models \Vel_{\stateb\in\dom(M^\ser)} \delta_{(M,\stateb)}$ (because $\stateb_2$ is one of those $\stateb$) we have that $(\stateb_2,\stateb'_2) \in \dom(M\otimes\umodel'')$. From that, $(\stateb_1,\stateb_2)\in R(\agent)$, and $(\stateb'_1,\stateb'_2) \in R'(\agent)$, follows that $((\stateb_1, \stateb'_1), (\stateb_2, \stateb'_2)) \in R^\otimes(\agent)$. By definition of ${\mathfrak R}$ we also have $((\stateb_2,\stateb'_2), \stateb'_2) \in {\mathfrak R}$.

\end{proof}
Note that we keep the states outside the serial submodel $M^\ser$ out of the bisimulation. Without the seriality constraint the `back' condition of the bisimilarity cannot be shown: given a $((\stateb_1,\stateb'_1), \stateb'_1) \in {\mathfrak R}$ and $(\stateb'_1,\stateb'_2) \in R'(\agent)$, but where $\stateb_1$ has no outgoing arrow for $\agent$, the required $\agent$-accessible pair from $(\stateb_1,\stateb'_1)$ does not exist. A special case of Proposition \ref{arbit} is the Corollary already referred to during the initial two-step construction, that achieves even isomorphy:

\begin{corollary} \label{isom}
Given $(M,\state)$, $(M', \state')$, and $\umodel'_r$ as above. Assume that $M$ is a bisimulation contraction. Then $(M \otimes \umodel'_r) \cong M'$.
\end{corollary}

\begin{proof}
In this special case we have that $(\stateb,\stateb') \in \dom(M\otimes\umodel'_r)$ iff $(M,\stateb) \models \pre'(\stateb')$ iff $(M,\stateb) \models \delta_{(M,\state)}$ for the point $\state$ of $(M,\state)$. As the last is only the case when $\stateb=\state$ (as $M$ is a bisimulation contraction), we end up with a domain consisting of all pairs $(\state,\stateb')$ for all $\stateb'\in \States'$, a 1-1-correspondence. The bisimulation ${\mathfrak R}$ above becomes the isomorphism ${\mathfrak I}(\state, \stateb') = \stateb'$.
\end{proof}


A different wording of Proposition \ref{arbit} is that for arbitrary finite epistemic states $(M,\state)$ and $(M', \state')$ also satisfying the serial submodel constraint, there is an update $(\umodel,\event)$ transforming the first into the second. A final appealing way to formulate this result is:
\begin{corollary} \label{corolbel}
Given are a finite epistemic state $(M,\state)$ and a satisfiable formula $\phi$. If all agents occurring in $\phi$ have non-trivial beliefs in state $\state$ of $M$, then there is an update {\em realizing} $\phi$, i.e., there is a $(\umodel,\event)$ such that $(M,\state) \models \dia{\umodel,\event} \phi$.
\end{corollary}

Using completeness of the logic, this further implies that all consistent formulas can be realized in any given finite model. We find this result both weak and strong: it is strong because any conceivable (i.e., using the same propositional letters and set of agents) formal specification can be made true whatever the initial information state. At the same time, it is weak: the current information state does apparently not give any constraints on future developments of the system, or, in the opposite direction, any clue on the sequence of events resulting in it; the ability to change the value of atomic propositions arbitrarily gives too much freedom. Of course, if one restricts the events to {\em specific protocols} (such as legal game moves \cite{hvd.pit:2006}, and for a more general treatment see \cite{jfaketal.TARK:2007}), the amount of change is constrained.

\paragraph{AGM belief revision and belief update} Our results on arbitrary belief change seem related to the postulate of success in AGM belief revision \cite{agm:1985}. AGM belief revision corresponds to epistemic change, and AGM (in their terminology) belief update \cite{katsuno:91a} corresponds to ontic change (unfortunately, in the AGM community `update' means something far more specific than what we mean by that term). Given this correspondence we can achieve only expansion by epistemic change, and not proper revision; and the combination of ontic and epistemic change can be seen as a way to make belief update result in belief revision. Apart from this obvious interpretation of epistemic and ontic change, one can also view our result that all consistent formulas can be realized, differently: a consequence of this is that for arbitrary consistent $\phi$ and $\psi$ there is an update $(\umodel,\event)$ such that $[\agent] \phi \imp \dia{\umodel,\event} [\agent] \psi$. In AGM terms: if $\phi$ is believed, then there is a way to revise that into belief of $\psi$, regardless of whether $\phi\et\psi$ is consistent or not. In other words: revision with $\psi$ is always successful. That suggests that our way of achieving that result by combining epistemic and ontic change might somehow simulate standard AGM belief revision. Unfortunately it immediately clear that we allow far to much freedom for the other AGM postulates to be fullfilled. It is clearly not a {\em minimal} change, for example. So walking further down this road seems infeasible.

\subsection{Postconditions true and false only}

The postconditions for propositional atoms can be entirely simulated by the postconditions true or false for propositional atoms. For a simple example, the public assignment $\atom := \phi$ can be simulated by a two-point update $\event$||$\Agents$||$\eventb$ (i.e., a nondeterministic event where all agents in $\Agents$ cannot distinguish $\event$ from $\eventb$) such that $\pre(\event) = \phi$, $\post(\event)(\atom) = \T$, $\pre(\eventb) = \neg \phi$, $\post(\eventb)(\atom) = \F$. In the public assignment $(\atom :=\phi, \atomb := \psi)$ to two atoms $\atom$ and $\atomb$ we would need a {\em four}-point update to simulate it, to distinguish all {\em four} ways to combine the values of two independent atoms.

The general construction consists of doing likewise in every event $\event$ of an update model. For each $\event$ we make as many copies as the cardinality of the powerset of the range of the postcondition associated with that event. Below, the set $\{0,1\}^{\dom(\post(\event))}$ represents that powerset.

\begin{definition}[Update model $\umodel^\truefalse$]
Given is update model $\umodel = (\Events, \arel, \pre, \post)$. Then $\umodel^\truefalse \isdef (\Events^\truefalse, \arel^\truefalse, \pre^\truefalse, \post^\truefalse)$ is a {\em normal update model} with
\begin{itemize}
\item $\Events^\truefalse=\bigcup_{\event\in\Events} \{ (\event,f) \suchthat f \in \{0,1\}^{\dom(\post(\event))}\}$
\item $((\event, f),(\event',f')) \in \arel^\truefalse(\agent)$ iff $(\event,\event') \in \arel(\agent)$
\item $\pre^\truefalse(\event,f) = \pre(\event) \et \Et_{f(\atom)=1} \post(\event)(\atom) \et
\Et_{f(\atom)=0} \neg \post(\event)(\atom)$
\item $\post^\truefalse(\event,f)(\atom) = \left\{ \begin{array}{ll} \T &\text{if }f(\atom)=1\\
                                                \F &\text{if }f(\atom)=0\\
                  \end{array} \right.
 $
\end{itemize}
\end{definition}

\begin{proposition} \label{unkelepup}
Given an epistemic model $M = (\States, R, V)$ and an update model $\umodel = (\Events, \arel, \pre, \post)$ with normal update model $\umodel^\truefalse$ defined as above. Then $(M \otimes \umodel) \bisim (M \otimes \umodel^\truefalse)$.
\end{proposition}
\begin{proof}
We show that the relation ${\mathfrak R}: (M \otimes \umodel) \bisim (M \otimes \umodel^\truefalse)$ defined as
\[
 ((\state,\event), (\state,\event,f)) \in {\mathfrak R} \text{ iff } (M,\state) \models \pre^\truefalse(\event,f)
\]
is a bisimulation. Below, the accessibility relations in $(M \otimes \umodel)$ and $(M \otimes \umodel^\truefalse)$ are also written as $R(\agent)$.

\begin{description}
\item {\bf atoms} \\
Let $(\state,\event,f)$ be a state in the domain of $(M \otimes \umodel^\truefalse)$. We have to show that for all atoms $\atom$, $(M,\state) \models \post(\event)(\atom) \eq \post^\truefalse(\event,f)(\atom)$. From the definition of $\post^\truefalse$ it follows that \[ \post^\truefalse(\event,f)(\atom) \Iff f(\atom)=1 \ . \] From $(M,\state) \models \pre^\truefalse(\event,f)$ and the definition of $\pre^\truefalse$ follows that \[ (M,\state) \models \post(\event)(\atom) \Iff f(\atom)=1 \ . \] Therefore  \[ (M,\state) \models \post(\event)(\atom) \eq   \post^\truefalse(\event,f)(\atom) \ . \]

\item {\bf forth} \\ Assume that $((\state,\event), (\state',\event')) \in R(\agent)$ and that $((\state,\event), (\state,\event,f)) \in {\mathfrak R}$. Let $f': \dom(\post(\event')) \rightarrow
  \{0,1\}$ be the function such that
  \[
   f'(\atom) = \left\{
   \begin{array}{ll}
    1 & \text{if } (M,\state')\models \post(\event')(\atom) \\
    0 & \text{otherwise } 
   \end{array}\right.
   \]
   Therefore $(M,\state') \models \pre^\truefalse(\event',f')$. Therefore
   $((\state',\event'), (\state',\event',f')) \in {\mathfrak R}$. From $((\state,\event), (\state',\event')) \in R^\truefalse(\agent)$ follows $(\state,\state') \in  R(\agent)$ and $(\event, \event') \in \arel(\agent)$. From $(\event, \event') \in \arel(\agent)$ and the definition of access $R^\truefalse$ follows $((\event,f), (\event',f')) \in R^\truefalse(\agent)$. From $(\state,\state') \in  R(\agent)$ and $((\event, f), (\event',f')) \in R^\truefalse(\agent)$ follows $((\state, \event,f),(\state',\event',f')) \in R(\agent)$.

\item {\bf back} \\ Suppose $((\state,\event), (\state,\event,f)) \in {\mathfrak R}$ and $(\state,\event,f),
   (\state',\event',f') \in R(\agent)$. From the last follows $(\state, \state') \in  R(\agent)$ and $((\event,f),(\event',f')) \in \arel^\truefalse(\agent)$, therefore also $(\event, \event') \in \arel(\agent)$. Therefore $((\state,\event), (\state,\event')) \in R(\agent)$. That $((\state',\event'), (\state',\event',f')) \in {\mathfrak R}$, is established as in {\bf forth}.
\end{description}
\end{proof}

\begin{corollary} \label{znollikjes}
The logic of change with postconditions true and false only is equally expressive as the logic of change with arbitrary postconditions.
\end{corollary}
Although it is therefore possible to use postconditions true and false only, this is highly unpractical in modelling actual situations: the descriptions of updates become cumbersome and lengthy, and lack intuitive appeal.

A transformation result similar to that in Proposition \ref{unkelepup} can {\em not} be established for the the logic with only singleton update models, i.e., the logic of public announcements and public assignments (as in \cite{kooi.jancl:2007}). If public assignments could only be to true and to false, then updates with assignments always result in models wherein the assigned atoms are true {\em throughout} the model, or false {\em throughout} the model. Therefore, there is no transformation of, e.g., $\underline{\atom}$||$\neg\atom$ into $\atom$||$\underline{\neg\atom}$ using public assignments and public announcements only. The construction above results in a {\em two-event} update model, that is not a singleton.

A transformation result as in Proposition \ref{unkelepup} immediately gives an expressivity result as in Corollary \ref{znollikjes} for the languages concerned. It is also tempting to see such a transformation result as a different kind of expressivity result. In two-sorted languages such as the one we consider in this paper one can then distinguish between the expressivity of two kinds of syntactic objects. A formula ($\phi$) corresponds to a class of models that satisfy that formula, and a modality ($\alpha$) corresponds to a relation on the class of models. The ability to express more relations does not necessarily lead to the ability to express more classes of models, nor vice versa. For example, concerning formulas, epistemic logic without common knowledge is equally expressive as public announcement logic without common knowledge, and even equally expressive as public announcement and public assignment logic without common knowledge \cite{kooi.jancl:2007}. But, as we have now seen, more and more relations between the models can be expressed.

\subsection{Single assignments only}

Consider update model $\event$||$\agent$||$\eventb$ for a single agent $\agent$ and for two atoms $\atom_1$ and $\atom_2$ such that in $\event$, if $\phi_1$ then $\atom_1 := \phi_2$ and $\atom_2 := \phi_3$, and in $\eventb$, if $\phi_4$ then $\atom_1 := \phi_5$ and $\atom_2 := \phi_6$. Can we also do the assignments one by one? In other words, does this update correspond to a sequence of updates consisting of events $\eventc$ in which at most one atom is assigned a value: the cardinality of $\dom(\eventc)$ is at most 1. This is possible! {\em First} we `store' the value (in a given model $(M,\state)$ wherein this update is executed) of all preconditions and postconditions in fresh atomic variables, by public assignments. This can be in arbitrary order, so we do it in the order of the $\phi_i$. This is the sequence of six public assignments $\atomb_1 := \phi_1$, $\atomb_2 := \phi_2$, $\atomb_3 := \phi_3$, $\atomb_4 := \phi_4$, $\atomb_5 := \phi_5$, and $\atomb_6 := \phi_6$. Note that such public assignments do not change the structure of the underlying model. {\em Next} we execute the original update but without postconditions. This is $\event'$||$\agent$||$\eventb'$ with $\pre(\event')=\pre(\event)=\phi_1$ and $\pre(\eventb')=\pre(\eventb)=\phi_4$ and with $\post(\event')=\post(\eventb')=\mathsf{id}$. Note that $\atomb_1$ remains true whenever $\event'$ was executed, because $\atomb_1$ was set to be true whenever $\phi_1$ was true, the precondition of both $\event$ and $\event'$. Similarly, $\atomb_4$ remains true whenever $\eventb'$ was executed. We have now arrived at the final structure of the model, just not at the proper valuations of atoms.

Finally, the postconditions are set to their required value, {\em conditional} to the execution of the event with which they are associated. Agent $\agent$ must not be aware of those conditions (the agent cannot distinguish between $\event'$ and $\eventb'$). Therefore we cannot model this as a public action. The way out of our predicament is a number of two-event update models, namely one for each postcondition of each event in the original update. One of these two events has as its precondition the fresh atom associated with an event in the original update, and the other event its negation, and agent $\agent$ cannot distinguish between both. The four required updates are
\begin{itemize}
\item $\event_1$||$\agent$||$\event'_1$ \ \ with in $\event_1$, if $\atomb_1$ then $\atom_2 := \atomb_2$ and in $\event'_1$, if $\neg\atomb_1$ then $\mathsf{id}$
\item $\event_2$||$\agent$||$\event'_2$ \ \ with in $\event_2$, if $\atomb_1$ then $\atom_3 := \atomb_3$ and in $\event'_2$, if $\neg\atomb_1$ then $\mathsf{id}$
\item $\event_3$||$\agent$||$\event'_3$ \ \ with in $\event_3$, if $\atomb_4$ then $\atom_5 := \atomb_5$ and in $\event'_3$, if $\neg\atomb_4$ then $\mathsf{id}$
\item $\event_4$||$\agent$||$\event'_4$ \ \ with in $\event_4$, if $\atomb_4$ then $\atom_6 := \atomb_6$ and in $\event'_4$, if $\neg\atomb_4$ then $\mathsf{id}$
\end{itemize}
Now, we are done. These four final updates do not change the structure of the model, when executed. Therefore, now having set the postconditions right, the composition of all these constructs is {\em isomorphic} to the original update model! The general construction is very much as in this simple example.

\begin{definition}[Update model $\umodel^\single$]
Given an update model $\umodel = (\Events,\arel,\pre,\post)$, update model $\umodel^\single$ is the composition of the following update models: {\em First} perform $\Sigma_{\event\in\Events} |\dom(\post(\event))+1|$ public assignments for fresh variables $\atomb,\dots$, namely for each $\event\in\Events$, $\atomb^\event_0 := \pre(\event)$, and for all $\atom_1,\dots,\atom_n \in \dom(\post(\event))$, $\atomb^\event_1 := \post(\event)(\atom_1)$, ..., $\atomb^\event_n := \post(\event)(\atom_n)$. {\em Then} execute $\umodel$ but with identity (`trivial') postconditions, i.e., execute $\umodel' = (\Events,\arel,\pre,\post')$ with $\post'(\event) = \mathsf{id}$ for all $\event\in\Events$. {\em Finally}, execute $\Sigma_{\event\in\Events} |\dom(\post(\event))|$ two-event update models with universal access for all agents wherein for each event just one of its postconditions is set to its required value, by way of the auxiliary atoms. For example, for $\event\in\Events$ as above the first executed update is $\event_1$||$\Agents$||$\event_2$ with in $\event_1$, if $\atomb^\event_0$, then $\atom_1 := \atomb^\event_1$, and in $\event_2$, if $\neg \atomb^\event_0$ then $\mathsf{id}$.
\end{definition}

Without proof the following proposition will be clear:

\begin{proposition}
Given epistemic model $M$ and update model $\umodel$ executable in $M$. Then $\umodel^\single$ is isomorphic to $\umodel$, and $(M \otimes \umodel^\single)$ is isomorphic to $(M \otimes \umodel)$.
\end{proposition}
This result brings our logic closer to the proposals in \cite{baltagetal:1999,hvdetal.aamas:2005} wherein only one atom is simultaneously assigned a value. The relation to other proposals will be discussed in Section \ref{other}.


\section{Comparison to other approaches} \label{other}

\paragraph*{Action model logic}
Dynamic modal operators for ontic change, in addition to similar operators for epistemic change, have been suggested in various recent publications. As far as we know it was first mentioned by Baltag, Moss, and Solecki as a possible extension to their {\em action model logic} (for epistemic change), in \cite{baltagetal:1999}. This was by example only and without a language or a logic. A precise quotation of all these authors say on ontic change may be in order:
\begin{quote} {\em Our second extension concerns the move from actions as
    we have been working them to actions which change the truth values of
    atomic sentences. If we make this move, then the axiom of Atomic
    Permanence\footnote{I.e.: $[\alpha]p \eq (PRE(\alpha) \imp p)$
      \cite[p.15]{baltagetal:1999}} is no longer sound. However, it is easy
    to formulate the relevant axioms. For example, if we have an action
    $\alpha$ which effects the change $p := p \et \neg q$, then we would
    take an axiom $[\alpha]p \eq (PRE(\alpha) \imp p \et \neg q)$. Having
    made these changes, all the rest of the work we have done goes
    through. In this way, we get a completeness theorem for this logic.}
  \cite[p.24]{baltagetal:1999} \end{quote} The logic that we present here
is a realization of their proposal, and we can confidently confirm that the
authors were correct in observing that ``all the rest (...) goes
through''. To obtain such theoretical results the notion of {\em
  simultaneous} postconditions (assignments) for a finite subset of atomic
propositional letters is essential; this feature is not present in
\cite{baltagetal:1999} (but introduced in \cite{vaneijck:2004}).

In a later proposal by Baltag \cite{baltag:2002} a fact changing action $flipP$ is proposed that changes (`flips') the truth value of an atom $P$, with accompanying axioms (for the proper correspondent action $\alpha$ resembling a single-pointed action model) $[\alpha]\atom \eq (\pre(\alpha) \imp \neg \atom)$ if ``$\atom$ changes value (flips) in $\alpha$'', and otherwise $[\alpha]\atom \eq (\pre(\alpha) \imp \atom)$ \cite[p.29]{baltag:2002}. The approach is restricted to ontic change where the truth value of atoms flips. In the concluding section of \cite{baltag:2002}, the author defers the relation of this proposal to a general logic of ontic and epistemic change to the future.

\paragraph*{Recent work in dynamic epistemics} More recently, in a MAS application-driven line of research \cite{hvdetal.aamas:2005,hvd.pit:2006} assignments are added to the relational action language of \cite{hvd.jolli:2002} but without providing an axiomatization. In this setting only change of {\em knowledge} is modelled and not change of belief, i.e., such actions describe transformation of $S5$ models only, such as Kripke models corresponding to interpreted systems. 

A line of research culminating in {\em Logics of communication and change}
\cite{jfaketal.lcc:2006} also combines epistemic and ontic change. It
provides a more expressive setting for logical dynamics than our
approach. The logic presented here is a sublogic of LCC. In
\cite{jfaketal.lcc:2006} the focus is on obtaining completeness via
so-called reduction axioms for dynamic epistemic logics, by extending the
basic modal system to PDL. Our treatment of postconditions, also called
substitutions, stems from \cite{vaneijck:2004}. In the current paper we
focus on specific semantic results, and, as said, we use a dynamic
epistemic `dialect', not full PDL.

A recent contribution on combining {\em public} ontic and epistemic change, including detailed expressivity results for different combinations of static and dynamic modalities, is found in \cite{kooi.jancl:2007}. Our work uses a similar approach to ontic events but describes more complex than public events: the full generality of arbitrarily complex events involves exchange of cards among subgroups of the public, and other events with a `private' (as opposed to public) character. 

Finally, a general dynamic modal logic is presented in \cite{renardel:cm}, where ontic changes are also studied. The semantics of this logic uses tree-like structures, and fixed points are introduced in the language to be able to reason about updates.

\paragraph*{Belief revision} An independent recent line of investigation combining epistemic with ontic change arises from the belief revision community. Modelling {\em belief revision}, i.e., epistemic change, by dynamic operators is an old idea going back to Van Benthem \cite{jfak.logcol:1989}. In fact, this is one of the two original publications---together with \cite{plaza:1989}---that starts the area of dynamic epistemic logic. For an overview of such matters see \cite{baltagetal.hpi:2007,hvd.dagstuhl:2005,hvdetal.del:2007}. But modelling ontic change---known as {\em belief update}  \cite{katsuno:91a}---in a similar, dynamic modal, way, including its interaction with epistemic change, is only a recent focus of ongoing research by Herzig and collaborators and other IRIT-based researchers \cite{herzigetal:2006,herzigetal:2005b,laverny:2006}. Their work sees the execution of an event as so-called {\em progression} of information, and reasoning from a final information state to a sequence of events realizing it as {\em regression}---the last obviously relates to planning. The focus of progression and regression is the change of the {\em theory} describing the information state, i.e.\ the set of all true, or believed, formulas. As already mentioned in Section \ref{semanticresults}, the results for arbitrary belief change in Proposition \ref{arbit} and following corollaries can potentially be applied to model belief update in the AGM tradition.

\paragraph*{Interpreted systems} 
In a way, dynamic epistemic logics that combine epistemic and ontic change reinvent results already obtained in the interpreted systems community by way of knowledge-based programs \cite{meyden.pricai:1997,faginetal:1995}: in that setting, ontic and epistemic change are integrated. Let us point out some correspondences and differences, using the setting of van der Meyden's \cite{meyden.pricai:1997}. This work investigates the implementation of knowledge-based programs. The transition induced by an update between epistemic states, in our approach, corresponds exactly to a step in a run in an interpreted system that is the implementation of a knowledge-based program; the relation between both is explicit in van der Meyden's notion of the {\em progression structure}. Now the dynamic epistemic approach is both more general and more restrictive than the interpreted systems approach. It is more restrictive because dynamic epistemics assumes perfect recall and synchronicity. This assumption is implicit: it is merely a consequence of taking a state transition induced by an update as primitive. But the dynamic epistemic approach is also somewhat more general: it does not assume that accessibility relations for agents are equivalence relations, as in interpreted systems. In other words, it can also be used to model other epistemic notions than knowledge, such as introspective belief and even weaker notions.

Knowledge-based programs consist of joint actions $\langle a_e, a_1,..., a_n \rangle$ where $a_e$ is an action of the environment and where $a_1, ..., a_n$ are {\em simultaneous} actions by the agents 1 to $n$. An agent $\agent$ acts according to a conditions of the form `{\em if $\phi'$ do $a'$, if $\phi''$ do $a''$, ...}' etc. Let us overlook the aspect that conditions $\phi'$ have the form of known formulas (by agent $\agent$). Still, such statements {\em look} familiar to our alternative format for what goes on in an event, as in `{\em for event $\eone$: if $\phi$, then $p_1 := \psi_1$, ..., and $p_n := \psi_n$.}' (see after Definition \ref{updatemodel} on page \pageref{updatemodel}). This correspondence is not really there, but the similar format is still revealing. The different cases in a knowledge-based program are like the different events in an update model, and they equally express non-determinism. This is a correspondence. There are also differences. In dynamic epistemics, the condition $\phi$ in  `{\em if $\phi$, then $p_1 := \psi_1$, ...}' is both an executability precondition {\em and} an observation. Inasmuch as it is an observation, it determines agent knowledge. In the interpreted systems approach, observations are (with of course reason) modelled as different from preconditions. The assignments such as $p_1 := \psi_1$ in the `then' part of event descriptions are merely the ontic part of that event, with the `if' part describing the epistemic part. Epistemic and ontic features {\em together} correspond to actions such as $a'$ in `{\em if $\phi'$ do $a'$}'. In the interpreted systems approach, epistemic and ontic features of information change are therefore not separately modelled, as in our approach.

\section{Further research}

An unresolved issue is whether updates can be described as compositions of purely epistemic events (preconditions only) and purely ontic events (postconditions only). In \cite{herzigetal:2006} it is shown for public events, for a different logical (more PDL-like) setting. Such a result would be in the line of our other normalization results, and facilitate comparison to related approaches.

The logic can be applied to describe cryptographic bit exchange protocols, including protocols where keys change hands or are being sent between agents. The logic is very suitable for the description of protocols for computationally unlimited agents, such as described in the cited \cite{fischeretal:1996,stiglic:2001}. Using dynamic logics may be an advantage given the availability of model checking tools for such logics, as e.g.\ the very versatile epistemic model checker DEMO \cite{vaneijck.demo:2004} by van Eijck. The current version of DEMO only allows epistemic events. But van Eijck and collaborators are in the process of extending DEMO with assignments (postconditions), needed to model events that are both epistemic and ontic. We eagerly await the completion of their efforts; our semantic results can then be demonstrated in that model checker.

The results for `arbitrary belief change' suggest yet another possibly promising direction. Under certain conditions arbitrary formulas are realizable. What formulas are still realizable if one restricts the events to those considered suitable for specific problem areas, such as forms of multi-agent planning? And given a desirable formula (a `postcondition' in another sense of the word), what are the initial conditions such that a sequence of events realizes it? This is the relation to AI problems concerning regression as pointed out in the introductory section \cite{herzigetal:2006}, and also to reasoning given specific protocols, such as always has been the emphasis for knowledge-based programs in the interpreted systems community \cite{faginetal:1995}, and as recently investigated in \cite{jfaketal.TARK:2007} in a dynamic epistemic context.


\bibliographystyle{plain}
\bibliography{biblio2007}

\end{document}